\documentstyle[12pt]{article}
\begin{document}
\begin{titlepage}
\begin{minipage}{13cm}
\begin{flushright}
\begin{tabular}{l}
HD--THEP--92--50\\[2pt]
TPI--MINN--92/62--T\\[2pt]
TUM--T31--32/92\\[2pt]
\end{tabular}
\end{flushright}
\begin{center}
{\LARGE Remarks on the Heavy-Quark Symmetry}\\[5pt]
{\LARGE in the Complex Plane}\\
\vspace{1cm}

{\large Patricia Ball}\\
{\large\em Physik-Department, TU M\"{u}nchen, D-8046 Garching, FRG}\\[2pt]
\medskip

{\large H.G.\ Dosch}\\
{\large\em Institut f\"{u}r Theoretische Physik, Universit\"{a}t Heidelberg,}\\
{\large\em D-6900 Heidelberg, FRG}\\[2pt]

\medskip

{\large M.A.\ Shifman}\\
{\large\em Theoretical Physics Institute, University of Minnesota,}\\
{\large\em Minneapolis, MN 55455, USA}\\[2pt]

\bigskip

{\large \today}\vspace{2cm}

{\bf Abstract}

\end{center}
{\small
\noindent Hadronic systems built from a heavy quark and a
cloud of light quarks
and massless gluons posess the Isgur-Wise symmetry resulting in  certain
relations  between various form factors. These relations can not be valid in
the entire complex plane. We identify contributions which unavoidably
violate strongly the symmetry  in the time-like domain near thresholds.
A possible
connection with the heavy-quark phenomenology is discussed in brief.
}

\vspace{2cm}

\begin{center}
{\em submitted to Physical Review D}
\end{center}
\end{minipage}
\end{titlepage}

\section{Introduction}
Strong interactions in quark-antiquark systems become simpler if the
mass of one of the quarks goes to infinity \cite{Shur1}. In particular, the
transition amplitudes of the type $(Q\bar q )\rightarrow (Q'\bar q )$,
where $Q$ is a generic notation for a heavy quark with mass $m_Q$,
which will be used throughout the paper, and $q$ denotes a light
(massless) quark, induced by the vector and axial vector currents $\bar Q'
\gamma_\mu Q$ and $\bar Q' \gamma_\mu \gamma_5 Q$, respectively,
are related by a symmetry which takes place even if
$m_Q\neq m_{Q'}$, but both masses, $m_Q$ and $m_{Q'}$, are parametrically
larger than $\Lambda_{QCD}$. This symmetry is
called the Isgur-Wise (IW) symmetry \cite{Isgu1}.
Moreover, the absolute normalization of the form factors in the small
velocity limit is fixed \cite{Volo1}.\footnote{Historically the observation
\cite{Volo1} that the transition form factors are predictable in the small
velocity limit for $m_Q\neq m_{Q'}$ served as an initial impetus for the
introduction of the IW symmetry \cite{Isgu1}.}

For finite quark masses the symmetry is, obviously, broken. The parameter
which governs the breaking of the IW symmetry can be represented
by some positive power of the quantity
\begin{equation}
\epsilon = \frac{M_{B^*}^2 -M_B^2}{M_B^2} \, .
\end{equation}
Here $B^*$ and $B$ denote generic vector and
pseudoscalar mesons (ground states) with the quark content $(Q \bar q )$.
The quadratic mass difference
\begin{equation}
\delta^2 = M_{B^*}^2 -M_B^2
\end{equation}
stays constant in the limit $m_Q\rightarrow\infty$; therefore, one
might think that
in this limit the heavy quark symmetry (essentially, independence of the
strong interactions of the spin orientation of the heavy quark) becomes
perfect. While in a large range of momentum transfer this should
be the case, there
exist kinematic domains where $\epsilon$ does not actually measure the
strength of the symmetry breaking. This happens, in particular, when the
form factors are considered in the time-like region, near the corresponding
thresholds. Under these circumstances the genuine symmetry breaking parameter
is $\delta^2/(q^2-4M^2)$ rather than $\epsilon$ ,
and relations stemming from the IW
symmetry do not hold. As a matter of fact, in certain instances the breaking
is even parametrically large. This simple observation will be quantitatively
worked out below.

This phenomenon is not unique, of course. Similar behavior is well-known,
say, for the isotopic invariance of strong interactions. Amplitudes which
are, generally speaking, believed to be isotopically invariant can exhibit
very strong deviations from the symmetry predictions provided that a
typical energy scale in the process at hand is not large compared to the
mass difference $m_d-m_u$. A classical example is the $D\bar D$
form factor in the vicinity of $\psi (3.77)$. The imaginary part of this
form factor in the resonance is strongly different in the channels $D^+D^-$
and $D^0\bar D^0$ due to the fact that the energy relase is about 40 MeV --
quite comparable to $2M(D^+)-2M(D^0)\approx 10$ MeV. It is quite typical
that the symmetry breaking effects, although large, are calculable in this
case.

In this note we analyse, in the same spirit, the peculiarities of the IW
symmetry breaking in the time-like region, mostly due to anomalous
thresholds \cite{Karp,Polk}. A  remark on possible phenomenological
implications in the so-called
molecular quarkonium (analog of the molecular charmonium
\cite{Deru1,Volo2}) is presented.
For definiteness we concentrate on the diagonal vector current
\begin{equation}
J_\mu = \bar Q\gamma_\mu Q,
\end{equation}
although the assertions made below are of a general nature and are
applicable, in principle, to other currents -- axial, non-diagonal, etc.
For the vector current one can consider three transition amplitudes:
\begin{equation}
<B|J_\mu |B>,\,\, <B|J_\mu |B^*>\,\,{\rm and}\,\, <B^*|J_\mu |B^*>.
\end{equation}
Taking into account current conservation one concludes that the first
and the second transitions are described by one form factor each,
$$ <B(p')|J_\mu|B(p)> = F_+ P_\mu , $$
\begin{equation}
<B(p')|J_\mu|B^*(p)> =-if\varepsilon_{\mu\alpha\beta\gamma}p'_\alpha
p_\beta\epsilon_\gamma ,
\end{equation}
while the third amplitude is represented by four form factors,
$$
<B^*(p')|J_\mu|B^*(p)> =
F_1(\epsilon\epsilon ')P_\mu +F_2[\epsilon_\mu (\epsilon 'P) +
\epsilon_\mu ' (\epsilon P)]
$$
\begin{equation}
+ F_3\frac{(\epsilon P)(\epsilon 'P)}{M_*^2}P_\mu
+F_4[\epsilon_\alpha(\epsilon 'P)
-\epsilon_\alpha '(\epsilon P)]\frac{q^2g_{\mu\alpha }-q_\mu
q_\alpha }{M_*^2}
,
\end{equation}
where
\begin{equation}
P=p'+p,\,\,q=p'-p,
\end{equation}
$M_*$ is the $B^*$ mass, $\epsilon$ and $\epsilon '$ are the polarization
vectors of the initial and final $B^*$'s. All six form factors,
$F_+,\,\, f,\,\, F_1,\,\, ...,\,\, F_4$, are functions of $q^2$.

The heavy quark symmetry reduces the number of independent functions in
the general decomposition (5) and (6) to one, in the limit
$m_Q\rightarrow\infty$ one obtains \cite{Isgu1} :
$$
\frac{1}{M}<B|J_\mu |B> = \xi (v'+v)_\mu,
$$
$$
\frac{1}{M}<B|J_\mu |B^*> =-i \xi \varepsilon_{\mu\alpha\beta\gamma}
v_\alpha ' v_\beta\epsilon_\gamma ,
$$
\begin{equation}
\frac{1}{M}<B^*|J_\mu |B^*> = \xi \{ -(\epsilon\epsilon ')(v'+v)_\mu
+[\epsilon_\mu (\epsilon '(v'+v)) +\epsilon_\mu '(\epsilon (v'+v))]\} .
\end{equation}
Here $v_\mu =p_\mu /M$, and the function $\xi$ depends in the IW limit only
on the product $y$ of four-velocities,
\begin{equation}
y=v'v.
\end{equation}
Moreover, $\xi (y=1) =1$ \cite{Volo1}. Let us draw the reader's attention
to the fact that $F_3$ and $F_4$ are predicted to vanish.

Any deviation from these relations will signal a violation of the IW
symmetry \footnote{The above reduction of six form factors to one for non-
relativistic heavy quarks, i.e.\ at $ |\vec v |\ll 1$, is known in the
literature for more than fifteen years \cite{Deru2}, see also Chapter 4
in the review paper \cite{Novi1}.}. An obvious violation occurs above the
threshold of the $B\bar B$ production but below $ B\bar B^*$. Indeed, in
this domain all form factors have imaginary parts associated with the
normal thresholds due to $B\bar B$. On the other hand, there is no
contribution to the imaginary part from $B\bar B^*$ and $B^*\bar B^*$.
It is quite clear that in the pseudoscalar $B$ meson the spin of the heavy
quark $Q$ is rigidly correlated with that of the light cloud. Hence,
the spin independence of the heavy quark interaction -- raison d'etre of
the IW symmetry and the origin of eq.\ (8) -- is totally lost.
In particular, the ``forbidden'' function $F_3$ appears. Below we will
demonstrate this assertion in a less trivial context of anomalous
thresholds. The existence of these thresholds is due to pion exchange.
A specific feature which makes their analysis interesting is
the fact that they can start parametrically much below the normal
thresholds, depending on the interplay between $\delta$ and $m_\pi$, the
pion mass. Certainly, in the real world both quantitites are such as they
are. It is instructive, however, to play with $m_\pi /\delta$  treating
this ratio as a free
parameter. One can easily change $m_\pi$ by adjusting the light quark mass
terms in the QCD lagrangian.
We will investigate different regimes allowed for the anomalous thresholds
in Sect.\ 2. Among other things it
will be seen that under certain conditions the anomalous
threshold starts at a value of $q^2$ which is
independent of the heavy quark mass. The corresponding contribution,
obviously, has no smooth IW limit in the
sense that its dependence on the momentum transfer does not reduce to a $y$
dependence, as prescribed by eq.\ (9). This particular contribution is not
leading at $m_Q\rightarrow\infty$. There are other regimes, however,
(which also defy
the heavy quark symmetry) where  we find
an enhancement by {\em positive} powers of $m_Q$ in the near-threshold
domain.

The issue of molecular quarkonium (bound states of the type $B^*\bar B^*$)
and its possible relation to the violations of the IW symmetry is
discussed in Sect.\ 3.

\section{Pion Exchange and Anomalous Thresholds}

First of all it is instructive to notice that the form factor of $B$ is
free from  anomalous thresholds  while that of the $B^*$ meson is
not. Indeed, let us consider the triangle graphs depicted in Figs.~1a, b.
Both graphs can lead to singularities on the physical sheet below the
normal thresholds the positions of which depend on the external masses.
The existence of such anomalous singularities has been first realized by
Karplus, Sommerfield and Wichman \cite{Karp}. From the standard Landau
theory \cite{Land1}
it is not difficult to find the conditions under which the
anomalous thresholds may occur.

Inspecting the dual diagrams associated with Figs. 1a, b we conclude the
following:

(i) The graph 1a has an anomalous threshold provided that
\begin{equation}
\frac{\delta^2}{M+M_*}\leq\mu\leq\delta .
\end{equation}
The beginning of the anomalous singularity is at
\begin{equation}
t_0^{BB} = 4M^2-\left( \frac{\delta^2-\mu^2}{\mu} \right)^2.
\end{equation}
To avoid numerous sub- and superscripts we have introduced the notations $t
\equiv q^2$, $\mu\equiv m_\pi$, $M\equiv M_B$ and $M_*\equiv M_{B^*}$.
The position of the anomalous threshold varies from zero at the lower
boundary of the interval (10) to the normal threshold
$4M^2$ at the upper boundary.

(ii) The condition on $\mu$ guaranteeing the existence of an
anomalous threshold in graph~1b is:
\begin{equation}
\frac{\delta^2}{M+M_*}\leq\mu\leq \frac{\delta}{\sqrt{1+(M/M_*)}} .
\end{equation}
The lower and upper boundaries in eq.\ (12) correspond to the position of the
anomalous threshold
at $M_*(M_*+M)$ and $(M+M_*)^2$, respectively.

The general expression for the position of the anomalous threshold in this
case is
\begin{equation}
t_0^{BB_*}=M^2+M_*^2+\frac{1}{2} (\delta^2-\mu^2) +
\left[ (\frac{4M_*^2}{\mu^2} -1)(\mu^2M^2 -
(\frac{\delta^2-\mu^2}{2})^2)\right]^{1/2} .
\end{equation}

A remarkable feature of eq.\ (11) is the fact the the position of the
anomalous threshold,
$t_0^{BB}$, needs not to scale as $M^2$ in the limit $m_Q\rightarrow\infty$;
if $\mu$ is chosen to lie close to $M_*-M$ other regimes are quite possible.
For instance, if $\mu = (M_*-M)(1+const.M^{-2})$ we find that $t_0^{BB}$
scales like $M^0$. The momentum transfer dependence then will be expressed
by  a function
of $(M^2/t_0^{BB})(1-y)$ rather than a function of $y$, and the IW scaling will
be obviously violated. Of course, the above regime is exotic since it
requires fine-tuning of the pion mass. One can always fix the pion mass and
then proceed to infinitely heavy quarks. This will push $t_0$ to $4M^2$
and will restore the $y$ dependence.

The position of the anomalous threshold in the complex plane, although
interesting by itself, says nothing about the relative weight of the
anomalous singularities. {\em A
priori} it is conceivable that they totally decouple at $m_Q\rightarrow\infty$.
Whether or not they actually
decouple depends on the behavior of the pion constant.

A brief reflection shows that the leading $M$ dependence of the pion
coupling to the heavy mesons  should be such that after proceeding to
non-relativistic normalization $M$ should disapper altogether
\cite{Volo2,Isgu2}. A relation between the $B^*B^*\pi$ and $B^*B\pi$
vertices can be
readily obtained by combining the heavy quark and chiral symmetries
\cite{Wise1,Burd1,Yan1}. In the relativistic normalization one gets
\begin{equation}\label{14}
B^*B\pi:\,\,\, \frac{4g}{f_\pi}M(\epsilon k) +O(M^0 ),
\end{equation}
\begin{equation}
B^*B^*\pi:\,\,\, -i\frac{2g}{f_\pi}
\varepsilon_{\alpha\beta\gamma\delta}
(p+p')_\alpha k_\beta \epsilon_\gamma ' \epsilon_\delta ,
\end{equation}
where $p$ , $p'$ and $k=p-p'$ are the momenta of the
initial and final heavy measons and the pion,
$g$ is a dimensionless coupling constant of order 1 \cite{Wise1,Burd1,Yan1},
and $f_\pi$ is the pion decay constant.
It is easy to check, for instance, that with eq.\ (\ref{14}) the
$D^*\rightarrow D\pi$ width does not contain the
heavy meson mass, in full accordance with the
arguments of refs. \cite{Volo2,Isgu2}.

We pause here to make an important remark. The fact that the $B^*B\pi$
vertex is proportional to $M$ (in the relativistic normalization) is a
consequence of the standard assumptions of the effective heavy quark
theory \cite{Eich1,Geor1}. Under certain kinematic conditions when the IW
symmetry is broken in the sense discussed in this note, the $BB^*\pi$
vertex may
 deviate from the standard scaling law exhibited in eq.\ (\ref{14}). In other
words, this vertex may be proportial to another power of $M$ for trivial
kinematic reasons which will become clear shortly. For the time being let
us stick, however, to eq.\ (\ref{14}), with $g=O(1)$.

It is important that the discontinuity of the amplitude at the anomalous
cut is determined by the triangular graphs  of Fig. 1 with all three
internal particles on mass shell \cite{Karp,Land1}. Therefore, the
pion coupling
constant which enters all our formul\ae\ is the on-mass-shell coupling.

Since our purpose is mostly illustrative it is reasonable to limit our
analysis to the anomalous singularities of graph 1a .The singularity of
this graph can, in principle, be arbirary close to $t=0$ while that of
Fig.\ 1b
necessarily lies higher than $2M^2$ (see eq.\ (13)) and, moreover, can be
pushed to the unphysical sheet provided that $\mu$ is chosen larger than
$\delta (1+(M/M_*))^{-1/2}$.

The contribution of the triangle graph depicted on Fig.\ 1a can be written
as follows:
\begin{equation}
{\cal A}_\mu = 16 i {\tilde g}^2\int \frac{d^4 k}{(2\pi )^4}
\frac{1}{k^2-\mu^2}\frac{(k\epsilon )(k\epsilon ') (p+p'+2k)_\mu }{((
p+k)^2-M^2)((p'+k)^2-M^2)}.
\end{equation}
where according to eq.\ (14):
\begin{equation}
\tilde g =\frac{gM}{f_\pi}.
\end{equation}

The anomalous cuts can be easily evaluated using Cutkosky \cite {Cutk}
rules,
yielding the imaginary parts $\rho_i(t)$ of the form factors $F_i(t)$,
\begin{eqnarray}
\rho_1(t) & = & \tilde{g}^2\frac{(4M_*^2\mu^2-\delta^4
-\mu^2(2\delta^2+\mu^2+t))(4M_*^2-2\delta^2-2\mu^2-t)}
{\sqrt{t} (4M_*^2-t)^{5/2}},\\
\rho_2(t) & = & - \tilde{g}^2\frac{(4M_*^2\mu^2-\delta^4
-\mu^2(2\delta^2+\mu^2+t))(2\delta^2+2\mu^2)}{\sqrt{t} (4M_*^2-t)^{5/2}},\\
\rho_3(t) & = & -2M_*^2\frac{\partial}{\partial t}\rho_1(t),\\
\rho_4 & = & 0.
\end{eqnarray}

It is quite obvious that the above expressions
badly violate the heavy-quark symmetry, since $\rho_1\neq -\rho_2$ and,
moreover, $\rho_3\neq 0$. The above equations should be supplemented by
the statement that the anomalous part does not vanish in the limit
$M\rightarrow \infty$. It is convenient to carry out the analysis of
the absolute normalization separately in two distinct cases:

(i) The  case in which $\delta $ and $\mu$ are fixed and thus the
anomalous threshold $t_0^{BB}$ is near the normal one. This is referred to
as the relativistic case;

(ii) the case in which $\delta$ stays fixed but $\mu$
decreases at least as $M^{-1}$ and thus the anomalous threshold is far
below the normal one. This is referred to as the non-relativistic case.

It should be noted that in the case considered here the anomalous
singularities are due to hadronic intermediate states and not to
quarks, as in the case of formfactors of heavy quarkonia, considered
in refs. \cite{Jaff1,Jaff2}.

\subsection{Relativistic case}

If $\delta$ and $\mu$ are fixed the anomalous threshold stays, according to
eq.\ (11), in the vicinity of the normal threshold, i.e.\ the relevant variable
is $u=4M_*^2-t$. Keeping only the leading in $M$ terms we get
\begin{eqnarray}
\rho_1(u) & = & \tilde g^2 \frac{(2 \delta^2 + 2\mu^2 -u)
(\delta^4 +2 \delta^2 \mu^2 +\mu^2(\mu^2-u))}
{ 2 M_* \, u^{5/2}} \\
\rho_2(u) & = & \tilde g^2 \frac{(\delta^2 + \mu^2)(\delta^4 +2
\delta^2 \mu^2 +\mu^2(\mu^2-u))}
{M_* \, u^{5/2}} \\
\rho_3(u) & = & - \tilde g^2  \frac{M_*^4}{2 M_*^{3/2} \,
u^{7/2}}\nonumber\\
      &   & \{ \mu ^2 u^2 - 3 u(\delta^4+4\, \delta^2 \mu ^2
+3\mu^4)+10(\delta^2+\mu ^2)^3\}
\end{eqnarray}

The anomalous cut in the $u$ plane extends from $(\delta^2+\mu^2)^2/\mu^2$
to $4\delta^2$. If $u=O(\delta^2)$ and $\tilde g = O(M)$ the discontinuities
are large rather than small,
\begin{equation}
\rho_1(u),\,\rho_2(u) \propto M,\,\, \rho_3(u)\propto M^3,
\end{equation}
leading to a huge violation of the heavy-quark symmetry in the
near-threshold domain (by the near-threshold domain we mean  an
interval of $t$ centered at $4M^2$ whose length is $O(\delta^2)$.)
Furthermore, it is quite trivial to do the dispersion integral with the
discontinuities given in eqs.\ (22-24). With
\begin{equation}
F_i^{an}\equiv \frac{1}{\pi} \int_{t_0}^{4M^2}\frac{\rho_i(t')dt'}{t'-t}
\end{equation}
one finds outside the near-threshod domain (at $t=O(M^2))$
\begin{equation}
F_1^{an}\sim F_2^{an}\sim \frac{1}{M},\,\, F_3^{an}\sim M .
\end{equation}
Specifically we obtain
\begin{equation}\label{28}
F_3^{an}(t=0) =\tilde g^2\frac{(\delta^2-\mu ^2)^3}{64 M_* \delta^5}
=g^2\frac{(\delta^2-\mu ^2)^3 M_*}{4 \delta^5 f_\pi^2}.
\end{equation}
The expression (\ref{28}), taken at its face value, would signal a
parametrically large
symmetry violation
even at $t=0$. We hasten to add that the anomalous terms we have
calculated should be (better to say, are expected to be) cancelled by
normal contributions of the hadronic graphs if one considers
the form factors $F_i$ outside the near-threshold
domain.

It is instructive to check how the above cancellation works by directly
computing the graphs of Fig.\ 1 (this is not a realistic computation of the
form factors for many reasons, of course, but just an exercise allowing one
to see the restoration of the symmetry). If $t$ is not especially close to
the threshold, the diagram 1a is not singled out. One must consider all
possible triangle graphs, both for $B^*$ and $B$, with the intermediate
states containing $BB$, $BB^*$ and $B^*B^*$. Adding them together, we
observe that all symmetry-violating terms with positive powers of $M$
cancel, if we start with formfactors satisfying eq.\ (8).

\subsection{Non-relativistic limit}

Let us consider another limiting case, when $\mu$ is only
slightly larger than $M_*-M$, so that $\mu$ scales as $M^{-1}$. Then the
anomalous threshold begins far below the normal one, i.e.\  $t_0\ll 4M^2$.

We start with
\begin{equation}
\mu=\frac{\delta^2}{2M\gamma }
\end{equation}
where $\gamma$ is a fixed number close to 1 but slightly less than one.
Then
\begin{equation}
t_0 \approx (1-\gamma^2)4M^2 .
\end{equation}
Neglecting $t$ compared to $M^2$ we get for the anomalous discontinuities
near the point $t=t_0$ the leading terms:
\begin{eqnarray}
\rho_1 (t) & = & \frac{ \tilde g^2\,\delta^4}{8 \gamma^2 M_*^3 \sqrt{t}}\,
(1- \gamma^2,)\\
\rho_2 (t) & = &- \frac{\tilde g^2 \, \delta^6}{16 \gamma^2  M_*^5\sqrt{t}}\,
(1- \gamma^2),\\
\rho_3 (t) & = & \frac{ \tilde g^2 \delta^4}{ 8 \gamma^2  M_* t^{3/2}}\,
(1- \gamma^2).
\end{eqnarray}

The anomalous contribution depends on $t$ through
$t/[4(1-\gamma^2)M_*^2]$, and the form factors scale in the following way
\begin{equation}
F_1^{an}\sim F_3^{an}\sim M_*^{-2},\,\, F_2^{an}\sim M_*^{-4}.
\end{equation}
In other words, the anomalous contribution decouples in the limit
$m_Q\rightarrow\infty$.

Of more interest is the case when $\gamma\rightarrow\ 1$ for
$M\rightarrow\infty$. Specifically, let us consider
\begin{equation}\label{scale}
\mu = (M_*-M)(1+\frac{x}{M^2}),
\end{equation}
where $x$ is a parameter of order $\delta^2$. Then
$t_0^{BB} \approx 8 x = O(M^0)$.
Under this choice -- when $M+\mu-M_*\equiv E =O(M^{-3})$ -- it is
reasonable to turn to a model according to which $B^*$ is a
non-relativistic bound state of $B\pi$, or, at least, such a four-quark
(molecular) component is present in $B^*$ with a certain prabability.

To make the situation more graphic, assume, at first, that $B^*$ completely
reduces to a loosely bound system of $B\pi$, analogous to the deuteron. Clearly
the spin symmetry between $B^*$ and $B$ is maximally violated in this
case, since the pseudoscalar meson, the would-be partner of the $B^*$, does not
look like this bound $B^*$ at all. Moreover, in this case one would expect
that the heavy-quark symmetry is violated not only in the near-threshold
domain, but everywhere in the complex plane.
The question is how this comes out formally.

The answer to this question is rather obvious. If $B^*$ is like the
deuteron, its coupling constant $g$ is rigidly fixed in terms of the binding
energy $E$ (see e.g.\ \cite{Land2} for an S-wave bound state)
and turns out to be much larger than that given in
eq.\ (17). Before we proceed to derive the  relation
between $E$ and $\tilde g$ let us
quote the expressions for the anomalous cuts in this limit,
\begin{eqnarray}
\rho_1(t)&=&\frac {- \tilde g^2 \delta^4}{32 M_*^5 \sqrt{t}}\,
(t-8 x ), \\
\rho_2(t)&=&\frac {\tilde g^2 \delta^6}{64 M_*^7 \sqrt{t}}\,
(t-8x ), \\
\rho_3(t)&=&\frac {\tilde g^2 \delta^4}{32 M_*^3\,t^{3/2}}\,
(t+8x).
\end{eqnarray}

If we had substituted $\tilde g =O(M)$ we would have got at $t=0$
\begin{equation}\label{scale2}
F_1^{an}(0)  \sim 1/M_*^3 \quad F_2^{an}(0) \sim 1/M_*^5
\quad F_3^{an}(0) \sim 1/M_*
\end{equation}

Now, let us derive the actual scaling law for $\tilde g$ in this scenario
of a loosely bound state.
Consider the non-relativistic contribution
 of
diagramm 1a , by doing the corresponding manipulations in
eq.\ (16). We put the pion on mass shell, i.e.\ take
only the imaginary part of its propagator, which just singles out the
anomalous contribution. We then replace the zero components of the
occuring Lorentz-vectors by their corresponding nonrelativistic
expressions:
\begin{equation}
p_0=M_* +{\vec p}\,^2/(2 M_*)\, , \quad k_0=\mu +{\vec k}\,^2/(2 \mu)
\end{equation}
The pion is nonrelativistic, since the binding energy and hence the virtual
momenta are small as compared to the pion mass according to (34). It is most
convenient to work in the Breit-frame, where $\vec p = -\vec p' =
\vec q /2 $ . We also neglect the pion mass $\mu$ against the heavy
mass $M_*$ . This yields for the zero component of the current:
\begin{equation}
\Delta^{nonrel} =   \frac{i \tilde g^2 \mu}{16 M_*(2\pi)^3} \int d^3k
             \frac{(\vec k \cdot \vec \epsilon - \mu/(2 M_*)\vec q
                   \cdot \vec \epsilon)
                   (\vec \epsilon' \cdot \vec k - \mu/(2 M_*)\vec \epsilon'
                    \cdot \vec q)}
                   {(\vec k-\mu/(2 M_*) \vec q)^2 + \alpha^2)
                   (\vec k+\mu/(2 M_*) \vec q)^2 + \alpha^2)}
\label{Dnr}
\end{equation}
where $\alpha = \sqrt{2 \mu E}$ with the binding energy
\begin{equation}
E = M + \mu - M_*
\end{equation}

We compare this expression to the form factor of the $B^*$-meson,
which is assumed to be a bound state of a $B$-meson and
a pion. We evaluate it in the impuls approximation
with the momentum space wave function
\begin{equation}
\vec \psi (\vec k) = N_{\alpha} f(k/\alpha) \vec k /(k^2+\alpha^2)
\label{wavef}\end{equation}
with $k= |\vec k|$ . The unphysical pole at $k^2 = - \alpha^2$
reflects the exponential tail of the pion cloud.

For the normalization constant we obtain
\begin{equation}
1/N_{\alpha}^2 =  4\pi \, \alpha \int \frac{f(y)^2 y^4}{(y^2+1)^2} dy,
\end{equation}
i.e.\ $N_{\alpha}^2 \sim C/\alpha$.
The impuls approximation for the zero component of the vector formfactor
yields:
\begin{equation}\label{Dia}
\Delta^{i.a.} =  \frac{ 2 M_* }{2 \pi^3} \int d^3 k \\
          \vec \epsilon' \cdot \vec \psi (\vec k-\mu/(2 M_*) \vec q)
               \quad \vec \psi^*(\vec k+\mu/(2 M_*) \vec q) \cdot \vec \epsilon
\end{equation}
A comparison of (\ref{Dnr}) and (\ref{Dia}) shows, that the function
$f(k/\alpha)$ in (\ref{wavef}) corresponds to some momentum cutoff in
(\ref{Dnr}), and that
the two expressions coincide, if the coupling constant is given by
\begin{equation}
\tilde g^2 = \frac{ C M_* ^2}{ \mu \alpha}
\end{equation}
where the constant $C$ depends on the momentum cutoff. Notice that the
relation between $\bar g$ and $E$ is different from that discussed in
\cite{Land2} due to the fact that the resonance we consider is P-wave.

Thus, for $\mu$ scaling like (\ref{scale}) we get
\begin{equation}
\tilde{g}^2 \propto M_*^5.
\end{equation}
It is rather straightforward to check that the form factors within this
scenario will have nothing to do with eq.\ (8), as expected.

A more realistic scenario will be to say that there is an admixture of the
$B\pi$ bound state in $B^*$. If the amplitude of this component scales as
$M^{-3/2}$, we have $\tilde g =O(M)$ and return back to eq.\ (\ref{scale2}).

The scale for the $t$-dependence of the anomalous contributions
is now however $t_0^{BB}$, which does not scale with $M_*$ and hence
we have for that case a --not very spectacular -- violation of
IW symmetry, which might be remarked as a spike near $t=0$ in
the derivatives of $F_i(t)$.

The question ``what is the actual admixture of the molecular $B\pi$ state in
$B^*$?'' is a dynamical one, and cannot be solved on the basis of
essentially kinematical arguments presented above.
The answer depends on the intricacies of the large distance dynamics and
can be a non-trivial function of $E$. While intuitively it is clear that
the overlap is less than 1, it needs not be as small as $M^{-3/2}$. Then
$\tilde g$
will not scale as prescribed by eq.\ (17).  If the overlap is larger
than $O(M^{-1/2})$, the spin symmetry violation in the entire complex
plane will survive in the limit $M\rightarrow\infty$.

We should mention that in the chiral limit $\mu = 0$ there is no
anomalous threshold on the physical sheet. In that case the $B^*$ is
unstable, but its width  vanishes as $ g^2 \delta^2/(f_\pi^2 M^3)$ .

\section{Molecular Quarkonium}

We would like to remind that the well-forgotten issue of the
molecular quarkonium \cite{Volo2} -- loosely bound states in the systems
$B\bar B,\,\, B\bar B^* +B^*\bar B,\,\, B^*\bar B^*$
is relevant to the discussion of the heavy-quark symmetry in the complex
plane. If the long-range forces binding these ``molecules''
are independent of the heavy quark spins \cite{Volo2},
i.e.\ eq.\ (8) holds, it is easy to check that the ratio of yields in these
channels is 1:4:7 \cite{Deru1}. This is the ratio stemming from eq.\ (8)
after squaring these expressions and averaging over the spacial
orientations of the momenta.
The question is how the symmetry is realized, if at all, in the ``molecular''
domain. (The ``molecules'' can be either bound states or resonances
above threshold.)

The answer depends on the relation between the spin splitting $\delta^2
/(2M)$ and the molecular-level splittings. For the lowest molecular levels
the latter are  expected to be parametrically larger than $\delta^2
/(2M)$. Then the spin symmetry will be reflected in an (approximate) triple
degeneracy of the levels. On average (summing over the triplet
of the degenerate levels) all symmetry relations, including 1:4:7 for the
ratio of yields, will be fullfilled, but at exactly the position
of each individual
resonance they are maximally violated. On the other hand,
if there are molecular states very close to the threshold,
$M_{res}-2M_*\leq \delta^2/(2M)$, it can well happen that such resonances
exist, say, only in the $B^*\bar B^*$ channel and are absent in $B\bar B$
(as is the case in charmonium). Then the form factors near the resonances
will not be symmetric.

\section{Concluding Remarks}

We have investigated the implications of anomalous thresholds
of heavy meson formfactors due to
pion exchange. For the actual values of the pion mass and the
quadratic mass splitting of the pseudoscalar and vector heavy mesons
we have found violations
of the IW symmetry near the production threshold of pairs of these
mesons. These violations are not only the trivial ones, due to the
finite quadratic mass splitting, but singularities of formfactors, which
are predicted to vanish become parametrically large, i.e.\ increase with
the heavy quark mass $m_Q$. The relation of these violations of the IW
symmetry near $t=4 m_Q^2$ with the molecular quarkonium is stressed.

If the pion mass is treated as free parameter, anomalous thresholds might
approach the region $t=0$ arbitrarily close. The quantitative importance
of such "low-lying" singularities depends on the degree of admixture of
the molecular $B\pi$  state in the $B^*$-meson.

\section{Acknowledgements}
This work has been partly done when one of the authors (M.S.) was visiting
the Institute for Theoretical Physics, Heidelberg, FRG, and CERN in summer
1992.
He is grateful to these theory groups for hospitality and financial
support. Useful discussions with M.\ Marinov, B.\ Stech, A.\
Vainshtein and M.\ Voloshin are gratefully acknowledged.

\newpage

\section*{Figures}

\setlength{\unitlength}{.2cm}

\begin{picture}(12,25)(0,0)
\multiput(.5,10)(2,0){5}{\oval(1,1)[t]}
\multiput(1.5,10)(2,0){5}{\oval(1,1)[b]}
\thicklines
\put(20,0){\line(0,1){5}}
\put(20,15){\line(0,1){5}}
\thinlines
\put(20,5){\line(-2,1){10}}
\put(10,10){\line(2,1){10}}
\multiput(20,5)(0,1){10}{\line(0,1){.5}}
\put(14,6){\makebox(0,0){$B$}}
\put(14,14){\makebox(0,0){$B$}}
\put(18,2){\makebox(0,0){$B^* $}}
\put(18,17){\makebox(0,0){$B^* $}}
\put(19,10){\makebox(0,0){$\pi $}}
\put(5,0){\makebox(0,0){\em Fig. 1a }}
\end{picture}
\begin{picture}(12,25)(-10,0)
\multiput(.5,10)(2,0){5}{\oval(1,1)[t]}
\multiput(1.5,10)(2,0){5}{\oval(1,1)[b]}
\thicklines
\put(20,0){\line(0,1){5}}
\put(20,15){\line(0,1){5}}
\put(10,10){\line(2,1){10}}
\thinlines
\put(20,5){\line(-2,1){10}}
\multiput(20,5)(0,1){10}{\line(0,1){.5}}
\put(14,6){\makebox(0,0){$B$}}
\put(14,14){\makebox(0,0){$B^*$}}
\put(18,2){\makebox(0,0){$B^* $}}
\put(18,17){\makebox(0,0){$B^* $}}
\put(19,10){\makebox(0,0){$\pi $}}
\put(5,0){\makebox(0,0){\em Fig. 1b }}
\end{picture}
\begin{picture}(12,25)(-20,0)
\multiput(.5,10)(2,0){5}{\oval(1,1)[t]}
\multiput(1.5,10)(2,0){5}{\oval(1,1)[b]}
\thicklines
\put(20,0){\line(0,1){5}}
\put(20,15){\line(0,1){5}}
\put(20,5){\line(-2,1){10}}
\put(10,10){\line(2,1){10}}
\thinlines
\multiput(20,5)(0,1){10}{\line(0,1){.5}}
\put(14,6){\makebox(0,0){$B^*$}}
\put(14,14){\makebox(0,0){$B^*$}}
\put(18,2){\makebox(0,0){$B^* $}}
\put(18,17){\makebox(0,0){$B^* $}}
\put(19,10){\makebox(0,0){$\pi $}}
\put(5,0){\makebox(0,0){\em Fig. 1c }}
\end{picture}

\bigskip

\bigskip

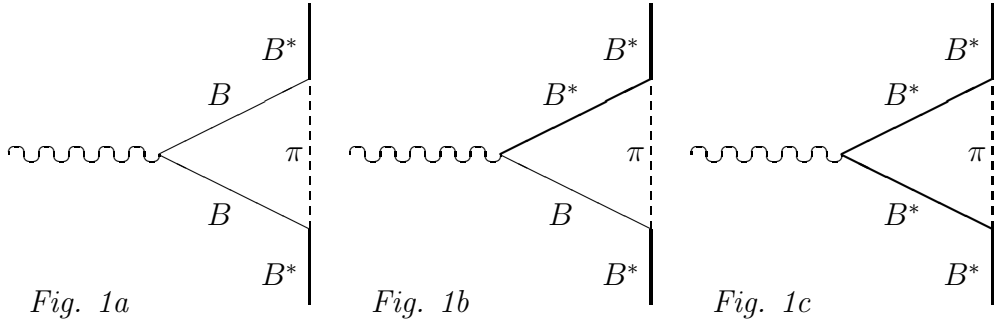
{\bf Figure 1}. Diagrams with pion exchange contributing to the
$B^*$ formfactor. Note that there is also a diagram corresponding to
Fig.\ 1b with the internal $B$- and $B^*$-lines interchanged.

\end{document}